\documentclass{article}
\usepackage{arxiv}
\usepackage[T1]{fontenc}
\usepackage[utf8]{inputenc}
\usepackage{graphicx}
\usepackage{threeparttable}
\usepackage{booktabs}
\usepackage{float}
\usepackage{array}
\usepackage{subfigure}
\usepackage{amsmath,amssymb,amsfonts}
\newcommand\dash{\nobreakdash-\hspace{0pt}}
\usepackage{algorithmic}
\usepackage{graphicx}
\usepackage{textcomp}
\usepackage{color,soul}
\usepackage{xspace}
\usepackage[square,numbers]{natbib}
\title{Deep Controllable Backlight Dimming}
\author{Lvyin Duan\thanks{ Equal contribution.}\\
Tianjin University\\%
University of Warwick%
\And Demetris Marnerides\footnotemark[1]\\%
University of Warwick%
\And Alan Chalmers\\%
University of Warwick%
\And Zhichun Lei\\%
Tianjin University%
\And Kurt Debattista\\%
University of Warwick%
}
\newcommand{\param}{p_a}
\newcommand{\tparam}{$\param$\xspace}
\begin{document}

\maketitle

\begin{abstract}
Dual-panel displays require local dimming algorithms in order to reproduce content with high fidelity and high dynamic range. In this work, a novel deep learning based local dimming method is proposed for rendering HDR images on dual-panel HDR displays.  The method uses a Convolutional Neural Network to predict backlight values, using as input the HDR image that is to be displayed. The model is designed and trained via a controllable power parameter that allows a user to trade off between power and quality. The proposed method is evaluated against six other methods on a test set of 105 HDR images, using a variety of quantitative quality metrics.  Results demonstrate improved display quality and better power consumption when using the proposed method compared to the best alternatives.
\end{abstract}

\section{Introduction}

High dynamic range (HDR) technology is capable of capturing,
storing and displaying a much wider dynamic range of luminance compared to the traditional standard or low
dynamic range (LDR) technologies. HDR imaging can significantly improve viewing
experiences and has been used in photography, gaming, films, medical and
industrial
imaging~\cite{debevec2008recovering}~\cite{marchessoux2016clinical}.

HDR is becoming one of the main features in display technology.
Seetzen et
al.~\cite{Seetzen:2004:HDR} developed the first LED\dash{based} HDR display
with a maximum luminance of approximately 8,500 $cd/m^2$ and a dynamic range of
50,000:1. This display is composed of two panels, a backlight panel and an LCD
panel, that are used for modulating the backlight luminance and maintaining
colour and details respectively. HDR displays of this kind, often termed dual\dash{panel} displays, are
capable of presenting a significantly higher luminance range compared to conventional
displays.

Backlight dimming (BLD) algorithms are designed for modulating the backlight of
dual\dash{panel} displays according to the displayed image content. To date,
many BLD algorithms have been proposed~\cite{narwaria2016dual}, mainly for LDR
images. In general, BLD algorithms can be divided into two
categories: global dimming and local dimming. Global dimming methods are mostly
used for small size LCD devices, such as smartphones and tablets. The
backlights of these devices are placed on the edges (edge\dash{lit}) because of
restrictions to their thickness. Local dimming algorithms are mostly used for the
devices which are directly backlit (direct\dash{lit}), such as TVs and computer
monitors.
Compared
with global dimming algorithms, local dimming algorithms are considered to perform better in
terms of image contrast and power
consumption~\cite{lai2008backlight}\cite{Xu2012}. Although local dimming can
also be used to control edge\dash{lit} devices, a number of areas can not be
controlled as effectively, unlike with directly back\dash{lit} devices.

Current methods are designed by display specialists and researchers using
hand-crafted features or utilising real\dash{time} optimisation, which can be
sub-optimal in the first case and may be time-consuming in the latter.  Recently,
data driven methods, in particular deep learning, have been used for a wide
range of applications in image processing due to their strong learning and
representation capabilities and efficiency. In particular, CNNs form
the basis for many current state of the art models in classification,
detection, image translation and synthesis~\cite{schmidhuber2015deep}. Deep
learning methods can bypass human expertise and heuristics by learning
directly from data. 

In this paper, a novel local dimming algorithm based on a CNN architecture is
proposed for displaying HDR images on dual\dash{panel} HDR monitors. The
proposed CNN can efficiently predict the backlight values for each dimming area
directly, providing a high fidelity reproduction of the original content. To
the best of our knowledge this is the first deep learning method proposed for
local dimming algorithms. Furthermore, the proposed method is conditioned via a
controllable power parameter that provides a trade-off between power
consumption and quality. 

The primary contributions of this work are: (a) the first learning based local
dimming method that uses a CNN model for rendering HDR images on a
dual\dash{panel} HDR display; (b) 
an adaptive optimisation procedure with an input-dependent adjustable loss
(c) a comprehensive objective evaluation of the
proposed algorithm against existing state\dash{of}\dash{the}\dash{art}
solutions.

\section{Background and Related Work}
A number of local dimming algorithms have been proposed to date. Furthermore,
CNNs have been extensively used for addressing problems of image processing.
This section introduces the basic structure of LC displays and presents an overview of existing local dimming algorithms as well as
relevant CNN based methods.

\subsection{Dual\dash{Panel} Display Technology}
Dual\dash{panel} displays consist of a high\dash{resolution} panel that
reproduces image details and colour, and a low\dash{resolution} backlight panel
that controls the contrast ratio. The high resolution panel corresponds to a
three\dash{channel} image $T$, while the low resolution backlight corresponds
to a set of $N$ values $\{B_k \mid k \in [1,N]\}$, placed on a single\dash{channel}
image, $B$. Each value corresponds to a coarse grained segment of the high
resolution image, according to the placement of the individual lights on the
backlight panel. For ease of notation, $B$ has the same resolution as $T$ but
all the values are zero except at the $N$ locations, $\mathcal{S}$, that
correspond to each of the backlight values $B_k$.

\begin{figure}[t]
	\centering
	\includegraphics[width=0.6\linewidth]{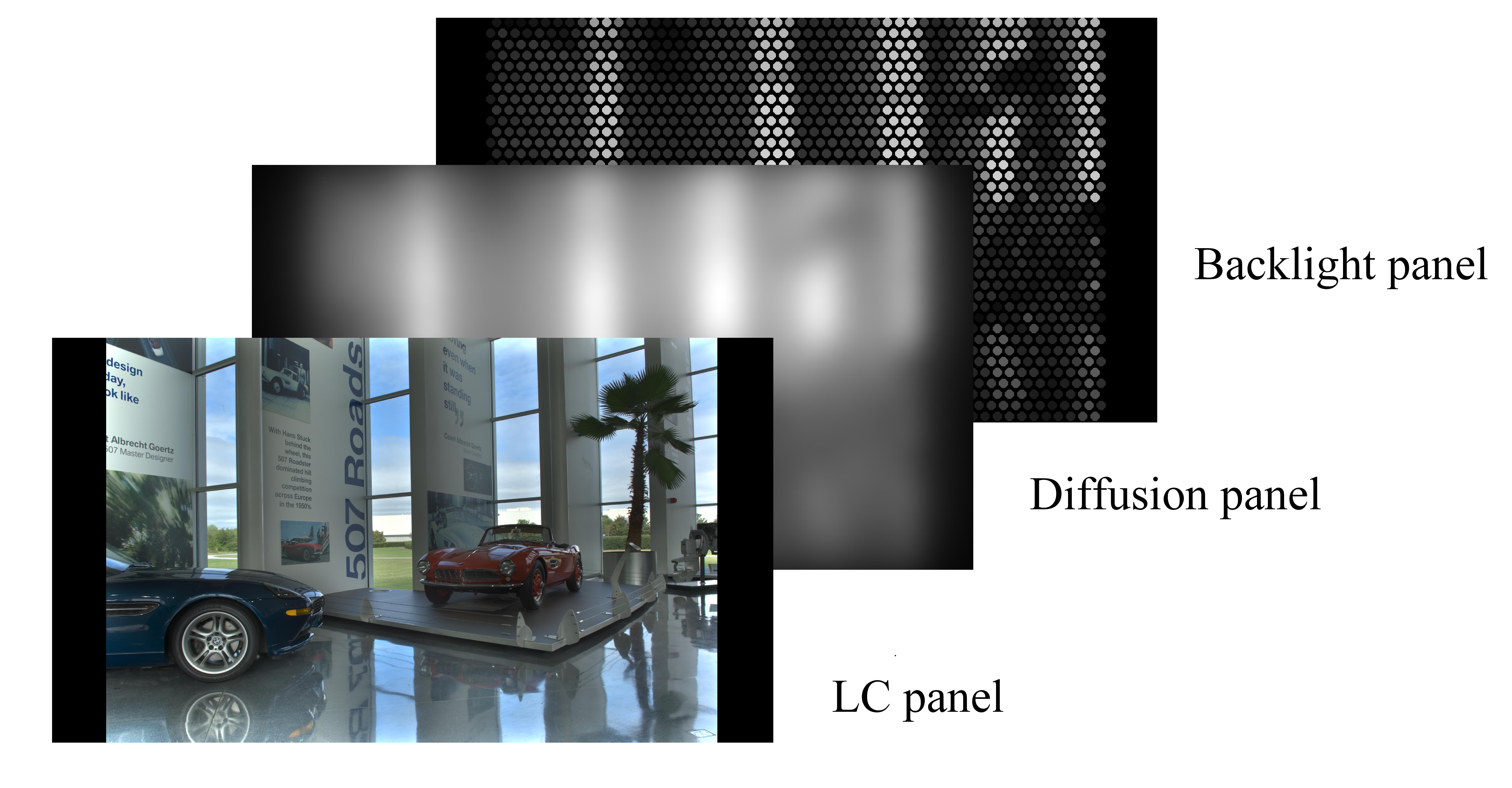}
	\caption{Structure of LC displays.}\label{fig:display_structure}
\end{figure}

Figure~\ref{fig:display_structure} shows the structure of dual panel LC
displays and their three main components: the backlight panel, the diffusion
panel and the LC panel. The backlight panel is the lighting source for the LC
panel, while the diffusion panel is used for smoothing and dispersing the
backlight in order to avoid huge luminance gaps and mismatch between
neighbouring pixels. The LC panel filters the backlight to create the three
channel image output at a high resolution.

\subsection{Existing Local Dimming Algorithms}

Local dimming algorithms can broadly be divided into three categories,
depending on their characteristics.

\subsubsection{Mathematical Statistics}

Statistics based local dimming algorithms obtain backlight values using straightforward
mathematical operators.

Funamoto et al.~\cite{funamoto} proposed the use of maximum and average
intensity of a given image segment. The maximum algorithm sets the intensity of
each backlight value to the maximum pixel value of the corresponding image segment.
The maximum approach is sensitive to noise, while the mean method tends to
produce excessively dim backlighting and can lead to significant clipping
artefacts.

\subsubsection{Local Image Characteristics}

BLD methods are based on assigning a backlight value that depend on each
local segment, rather than taking simple maximum or average values.

Cho and Kwon~\cite{cho2009backlight} proposed a BLD method to improve
image quality using a correction term to adjust the
average pixel intensity by considering the local difference between the
maximum and average luminance. In addition, a new method for reducing the
clipping artefacts of LCD images was used to preserve the image quality. A
similar method developed by Zhang et al.~\cite{zhang2012dynamic} who also computed
a correction term as the ratio of the difference of maximum and average
luminance to obtain the backlight values. Lin et al.~\cite{Liao2008Dynamic}
inversed the cumulative distribution function (from a global histogram) to map
a weighted mean of the maximum and average pixel values of each backlight
segment for the resulting backlight values.  Other methods, such as that
introduced by Nam~\cite{nam2011low}, consider both local and global brightness
in order to find a better trade\dash{off} between enhancing local contrast and
preserving the overall appearance of the LCD images. A roll\dash{off}
scheme was used to enhance image details in the high\dash{level} grey areas.
Cho et al.~\cite{cho2011two} used an image metric to obtain the intensity of
the backlight and refined these values by considering both local block lighting
and the lighting from neighbouring blocks.  Other BLDs were developed to
preserve the image quality, including Kang and Kim~\cite{kang2011multi} who  considered the
pixel distribution of an image using multiple histograms. Hsia
et al.~\cite{Chien2013High} proposed a method to improve the LCD image
resolution by enhancing the weak edges of each image segment.

\subsubsection{Optimal Methods} In BLD methods, clipping artefacts are the most
significant problem that effects the displayed image quality. To keep the
balance between displayed image and backlight values, some optimal BLD
algorithms have been proposed.

The BLD algorithm developed by Kim et al.~\cite{kim2009reduce} is based on a
decision rule: searching the optimal dimming value by comparing
the light\dash{leakage} measure and the clipping measure to keep the
light\dash{leakage} and clipping lower.   Shu et al.~\cite{shu2012optimal} approached
the local dimming of LED backlight LC displays as an optimisation problem to
obtain a higher visual quality. Zhang et al.~\cite{zhang2018optimal}
also proposed an optimal method to maintain a balance between LCD image
quality and power consumption. Cha et al.~\cite{cha2015optimized} presented an
efficient optimised BLD method for edge\dash{lit} lighting\dash{emitting} diode
backlight to reduce image quality fluctuation.  Another category of backlight
modulation methods, such as that proposed by e.g. Albrecht et
al.~\cite{albrecht2010sorted}, are
based on a point spread function (PSF) to exploit the knowledge of light
diffusion and model how light diffuses from a source.  There have also been
other approaches, such as those introduced by Burini et
al.~\cite{burini2012image} and Mantel et al.~\cite{mantel2013controlling},
which focus primarily on achieving a trade\dash{off} between clipping and
leakage. Forchhammer and Mantel~\cite{mantel2017viewpoint} extended the method
proposed by Mantel et al.~\cite{mantel2013controlling} further to multiple
viewers taking into account clipping and leakage as well as reflections of the ambient
light. To keep the LCD image quality, Seok\dash{Jeong} Song et.al~\cite{song2019deep}
proposed a pixel compensation algorithm based on deep learning for local
dimming algorithms on the quantum\dash{dot} display.

Although there have been many BLD algorithms developed for enhancing image
quality, these methods mostly target LDR images. To render HDR
images on dual\dash{panel} displays, Seetzen et al.~\cite{Seetzen:2004:HDR} created
a method to solve this problem by splitting HDR images into two layers using
square root of the image luminance channel. To assess the impact of HDR image
rendering on both subjective and objective scores, Zerman et
al.~\cite{zerman2015effects} proposed a method for HDR image rendering for the
SIM2 HDR47 display by minimising power consumption and
maximising the fidelity to the target pixel values. Narwaria et
al.~\cite{narwaria2016dual} also proposed an HDR image rendering solution
which used
a gradient\dash{based} optimisation to minimise the difference between the
theoretical backlight map and the computed light map.

Duan et al.~\cite{Duan} explored the relationship between LCD image quality and
backlight intensity and proposed an objective evaluation method for BLD
methods, and also conducted a subjective experiment to validate results. The
results demonstrated a strong correlation between objective and subjective
evaluation of different BLD algorithms.

\subsection{CNNs for Luminance Processing}
Recently, CNNs have been used for addressing a large range of problems related
to luminance processing because of their excellent performance and learning
capabilities for analysing image characteristics.

Yannick Hold\dash{Geoffroy} et al.~\cite{hold2017deep} presented a CNN based
technique to estimate high dynamic range outdoor illumination.  A number of
methods using CNNs have also been presented for Tone Mapping (HDR to LDR) and
Inverse Tone Mapping (LDR to
HDR)~\cite{eilertsen2017hdr,marnerides2018expandnet,li2018multi}.  To the best
of our knowledge, there are no local dimming methods using CNN architectures
for HDR images.

\begin{figure*}[t!]
	\centering
	\includegraphics[width=1.0\linewidth]{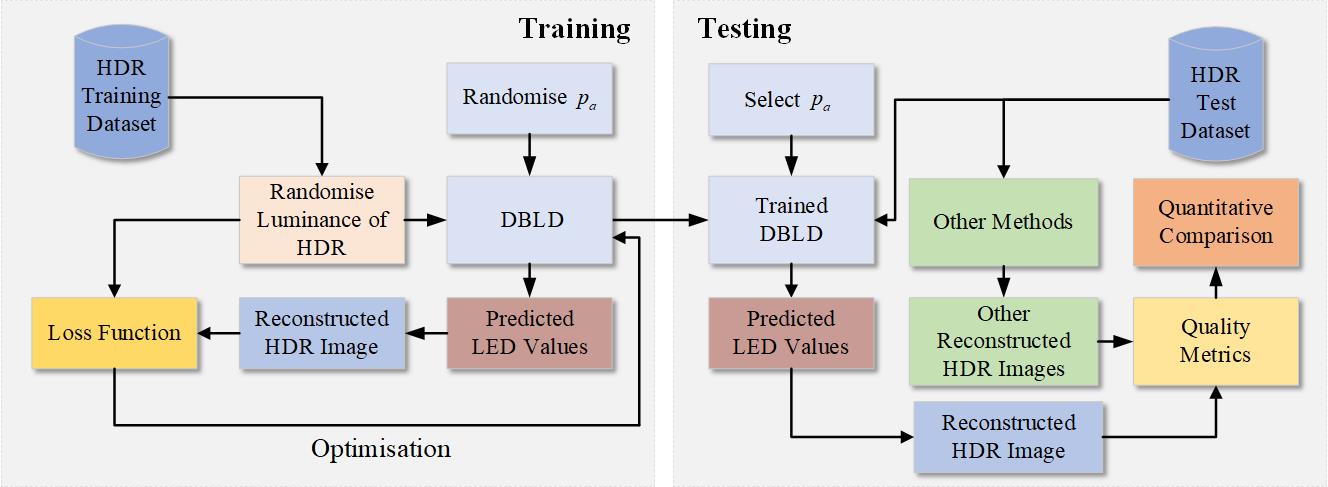}
	\caption{The framework for the training and testing of the proposed method.}\label{Framework}
\end{figure*}

\section{Method}

As discussed in the previous section, a variety of BLD algorithms have been proposed to date. More
importantly, most methods are based on modeller expertise~\cite{funamoto}, with
choices that can seem arbitrary and may not be optimal. Furthermore, 
non learning\dash{based} methods can ignore abstract and high level
image features that are deemed important in many imaging applications.

The proposed deep BLD method (DBLD) addresses these issues by using a parametric model to
process an input HDR image and directly predict the backlight values. The model
is optimised directly from data, avoiding modeller bias and heuristics.  The
parametric model of choice is a CNN, trained on a dataset of HDR images and
optimised to maximise the fidelity of the displayed HDR image and can be controlled via a power parameter, \tparam, that provides a balance between power consumption and quality. 

As shown in Figure \ref{Framework}, the procedure followed in this work is divided into two phases, a training phase
and a testing phase. In the training phase, the CNN is randomly initialised and
then optimised using an HDR dataset, by minimising a loss function. This is
performed only once and the optimised parameters are then used in the testing
phase to evaluate the method's performance by comparing it quantitatively with
other algorithms.  The method also makes use of \tparam to control how much power the LEDs consume. This is achieved via a novel loss function formulation that takes \tparam into account. 

\subsection{Network Architecture}

\begin{figure*}[t!]
	\centering
	\includegraphics[width=1.0\linewidth]{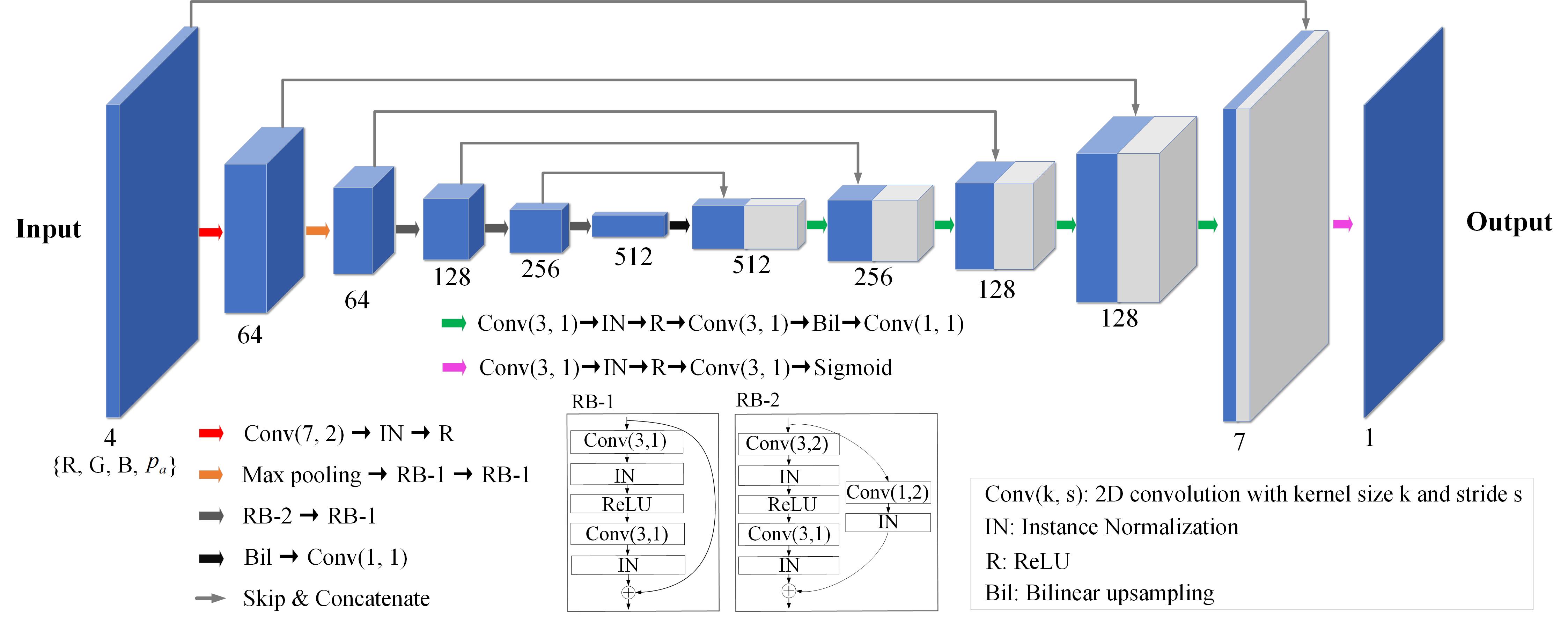}
    \caption{Diagram of the DBLD CNN architecture.}\label{fig:cnn}
\end{figure*}

The proposed architecture, shown in Figure~\ref{fig:cnn}, is based on the UNet
architecture~\cite{ronneberger2015u}, which is composed of two main parts, an
encoder and a decoder, both composed of multiple convolutional layers. The
encoder progressively downsamples the feature resolution until it reaches a low
resolution bottleneck, which is then progressively upsampled by the decoder. At
each resolution, features from the encoder are propagated directly to the
decoder and concatenated, effectively combining multiple scales and speeding up
convergence at optimisation. 

The encoder used is a residual network architecture~\cite{he2016deep} with 18
layers. Residual networks are formed from residual blocks, where the output of
the main computation of each block is added to its input, thus allowing better
gradient flow and improved training of deeper networks. The implementation is
taken directly from the ``resnet-18'' architecture in the PyTorch model
library~\cite{paszke2019pytorch}. The 18-layer resnet architecture is the most
lightweight of the commonly implemented residual networks. It downsamples five
times and uses $3\times3$ convolutions, except from the first layer which is of
size $7\times7$ and the residual-connection convolutions that are of size
$1\times1$ and are used to match the input-output feature sizes of each block
when they differ.

The decoder consists of five upampling layers that use bilinear upsampling
followed by blocks of \{$3\times3$ convolution - normalisation - activation -
$3\times3$ convolution\}, matching the feature sizes of the encoder at each
resolution. The ReLU activation~\cite{nair2010rectified} is used both in the
encoder and the decoder, along with Instance
Normalisation~\cite{ulyanov2016instance}, to help with convergence in the
optimisation. Instance Normalisation is preferred to the more commonly used
Batch Normalisation~\cite{ioffe2015batch} for small batch sizes in gradient
descent. In this work, the batch size consists of only one image at each
iteration due to GPU memory constraints, since training is performed on Full-HD
images. The model has a total of 13,782,031 parameters. Despite the large
number of parameters, processing is quick, since most of the computation is
performed on lower resolutions due to the use of the UNet architecture.

The network accepts a total of four channels of resolution 1,920 $\times$
1,080, consisting of the RGB channels of the HDR image, $I$, in the [0, 1]
range, along with a uniform single channel that holds the power parameter,
$\param \in [0, 1]$, which adapts the power consumption of the predicted
backlight values.  The output of the network, $\tilde{B} \in [0,1]$, is a
single channel image containing the backlight predictions at full resolution
and is the result of a logistic (sigmoid) function following the final
convolution. The final backlight prediction, $B$, is formed by selecting the
$N$ pixels corresponding to the $N$ LED lights in the backlight panel of the
SIM2 display. These are selected as the central pixels of the corresponding
areas of the image in $\tilde{B}$.

The final model for the backlight prediction, $B$, can be expressed as:

\begin{equation}
    B(I)_{i,j} = \begin{cases}
        f_{\text{CNN}}(I, \param)_{i,j}, & \text{if } (i,j) \in \mathcal{S},\\
        0, & \text{otherwise},
    \end{cases}
\end{equation}
\noindent where $\mathcal{S}$ is the set of centres
of the pixel neighbourhoods that correspond to the individual lights in the backlight
panel.

\subsection{HDR reconstruction}
\label{sec:hdr_reconstruction}

As shown in the work by Duan et al.~\cite{Duan}, the displayed HDR image can be
simulated and reconstructed artificially, for objective comparisons that agree with
subjective experiments. Hence, such a reconstruction can form a valid representation of
the displayed image, therefore allowing for its use as part of the loss function when
optimising the CNN. In theory, the resulting displayed image, $\tilde{I}$ is given by:

\begin{equation}\label{eq:displaymodel}
\tilde{I} = D\odot T,
\end{equation}

\noindent where $T$ is the transmittance of the LC panel, $D$ is the smoothened
backlight intensity from the diffusion panel. $\odot$ denotes the
(pixel\dash{wise}) Hadamard product operator, broadcasted channel\dash{wise}.
In general, the transmittance, $T$, is driven by the grey level of each pixel
from every colour channel of the LCD image, $C$.

The diffusion panel output, $D$, can be estimated from the backlight values
as the result of the convolution of the displayed backlight image,
$B$, with the PSF~\cite{forchhammer2016hdr}, $g$, of the diffusion panel:

\begin{equation}\label{eq:diffusionimage}
D = (g*B)_{i,j} = \sum_{x=1}^{W_{g}} \sum_{y=1}^{H_{g}} g_{x,y}B_{i-x, j-y},
\end{equation}

\noindent where $N$ denotes the total number of backlight values and $W_g$ and
$H_g$ are the width and height of the PSF filter respectively. $D$ is often referred to as the baseline luminance.

The loss function presented in Section~\ref{sec:loss} requires the
reconstructed HDR image, $\tilde{I}$, which in turn requires evaluation of the
baseline luminance, $D$. $D$ is estimated by convolving the backlight prediction,
$B$, with the PSF, $g$, following equation~\ref{eq:diffusionimage}. However,
the PSF for the modelled display is given as a single channel filter of size
$1,000 \times 1,000$. Fast differentiable convolution with large filters is not
directly implemented (at the time of writing) in modern deep learning
libraries~\cite{largekernelbug}. Most libraries optimise small convolutions,
e.g. with $3\times3$ kernels, since almost all CNN architectures use relatively
small kernels. Thus, the PSF convolution was implemented from scratch using
base (differentiable) PyTorch operations~\cite{paszke2019pytorch}.

In particular, the convolution is implemented using the convolution theorem,
applied on $B$ and $g$:

\begin{equation}
 D =  B * g = \mathcal{F}^{-1} \left( \mathcal{F}\left(B\right)\odot \mathcal{F}\left(g\right)  \right),
\end{equation}

\noindent where $\mathcal{F}$ is the Fourier Transform operator, in combination
with the Discrete Fourier Transform (FFT):

\begin{equation}
    S_{u,v} = \mathcal{F}\left(T\right) =  \frac{1}{\sqrt{HW}} \sum_{h=0}^{H-1}\sum_{w=0}^{W-1}
 T(h,w)e^{-2\pi i \left( \frac{hu}{H} + \frac{wv}{W}\right) },
\end{equation}

\noindent where T is the input in coordinate space and $S$ is the
representation of the input in fourier space. \(H\) and \(W\) are the height
and width of the image respectively. The Fourier transform is performed using
the Fast Fourier Transform (FFT) algorithm. This implementation for
convolutions with large kernels is much faster and uses less memory in contrast
to the default optimised convolution based on the cudnn library that would get stuck
and not complete the computation on the same machine~\cite{chetlur2014cudnn}.

\subsection{Loss Function}\label{sec:loss}

The loss function, $L$, consists of two parts, a smooth $L_1$
regression loss, $L_{\text{reg}}$, and an additional magnitude regularisation term,
$L_{\text{mag}}$, that also adapts power consumption by restricting the magnitude of the backlight predictions via the user-provided scalar power parameter, \tparam. The total loss is given by:

\begin{equation}
    L(\tilde{I}, I) = L_{\text{reg}}(\tilde{I}, I) + \param \beta L_{\text{mag}}(B),
\end{equation}

\noindent where $\tilde{I}$ is the HDR image reconstructed from the backlight
predictions of the model using the method described in
Section~\ref{sec:hdr_reconstruction} and $I$ is the target HDR image. $\beta$ is a
hyper-parameter adjusting the magnitude of the regression loss that helps with
levelling the gradient contribution of the two partial losses for improved
convergence.

The magnitude regularisation term, $L_{\text{mag}}$, is given by:

\begin{equation}\label{eq:magnitude_loss}
    L_{\text{mag}}(B) = \frac{1}{M_{\text{max}}}\sum_{(i,j)}B_{i,j},
\end{equation}

\noindent where $M_{\text{max}}$ is the maximum consumption, when all
backlights take their maximum value. The magnitude regularisation term
restricts power consumption by penalising large backlight values.  The
non-learned user-provided power parameter, \tparam, appears directly in the
loss function, changing the form of the loss during training by adjusting the
contribution of the magnitude term $L_{\text{mag}}$. Lower \tparam values allow
higher $L_{\text{mag}}$ values in the loss, thus allowing higher power
consumption.

\begin{figure*}[!ht]
\centering
\includegraphics[width=0.33\linewidth]{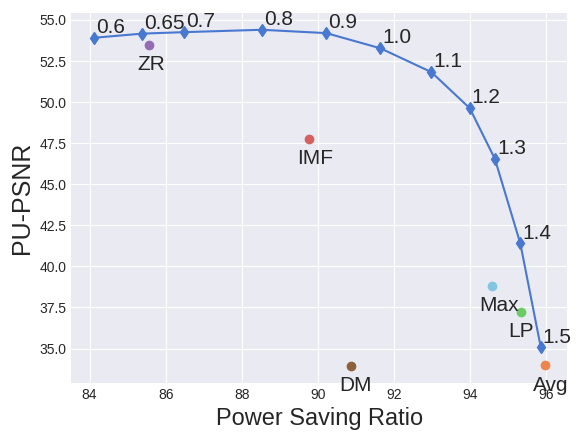}%
\includegraphics[width=0.33\linewidth]{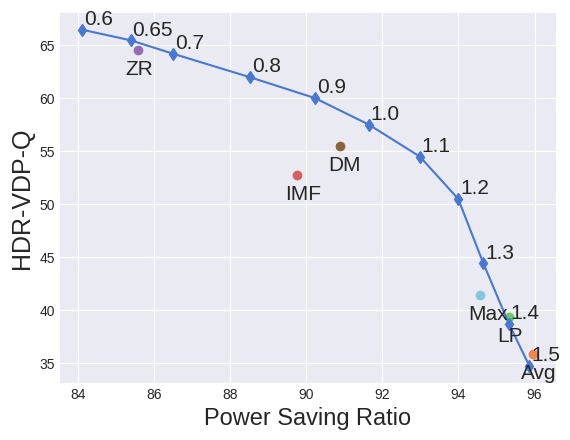}%
\includegraphics[width=0.33\linewidth]{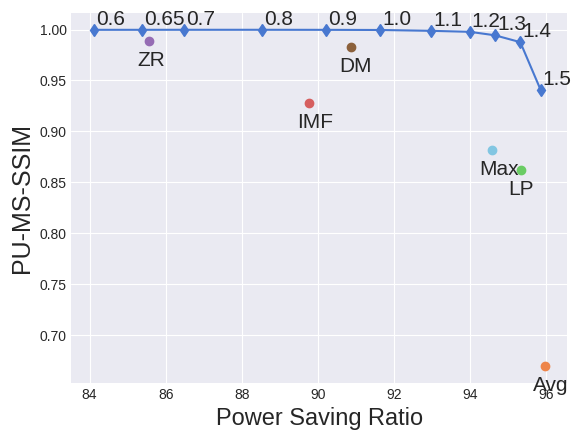}
\caption{Comparison of median values of PU-PSNR, HDR-VDP-2.2, and
PU-MS-SSIM against PSR. Adjusting \tparam allows for the proposed DBLD method (blue line) to adapt power consumption for improved quality.}\label{fig:plots}
\end{figure*}

\begin{figure*}[!t]
\centering
\includegraphics[width=0.45\linewidth]{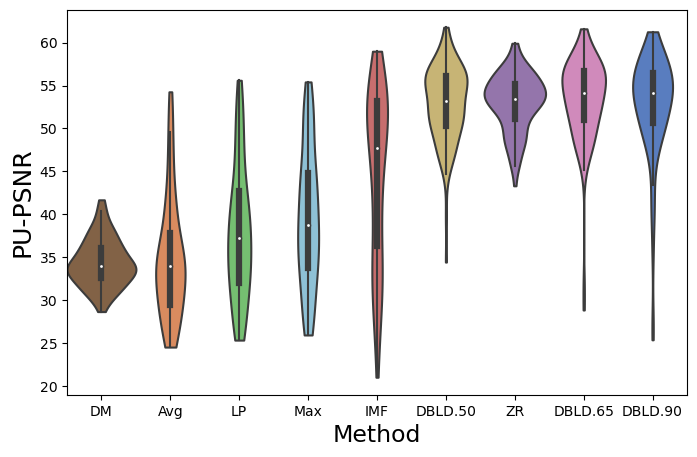}%
\includegraphics[width=0.45\linewidth]{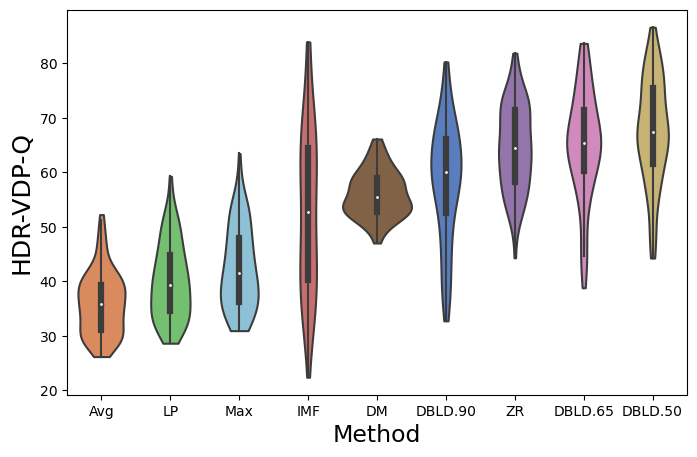}\\
\includegraphics[width=0.45\linewidth]{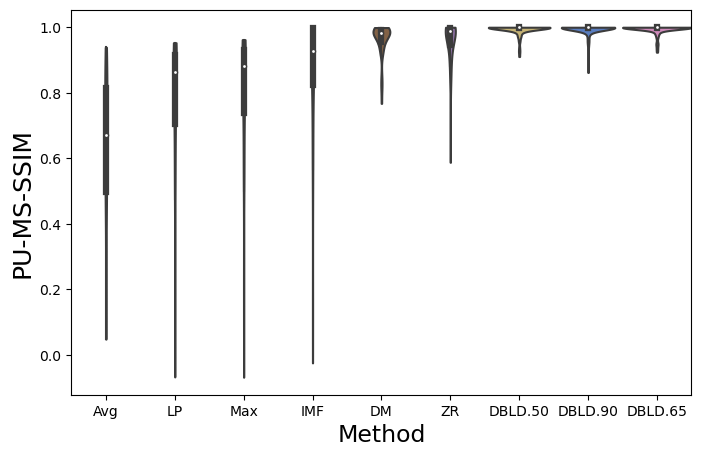}%
\includegraphics[width=0.45\linewidth]{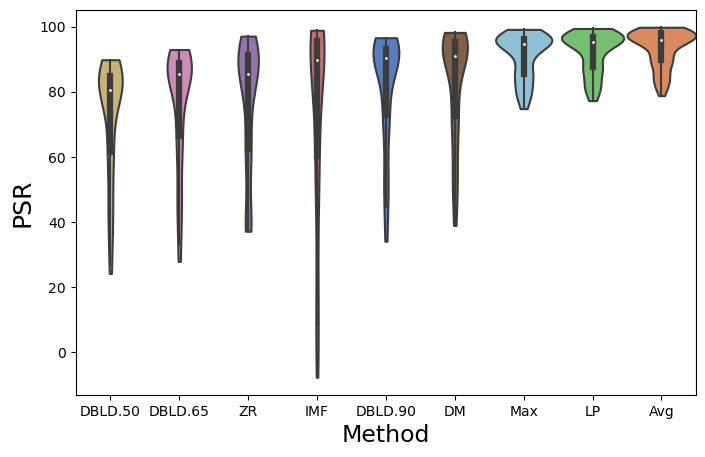}%
\caption{Comparison of the distributions of PU-PSNR, HDR-VDP-2.2, PU-MS-SSIM
and PSR for all methods. The proposed DBLD method is evaluated at different
values of \tparam (0.5, 0.65 and 0.9).}\label{fig:violinplots}
\end{figure*}

\subsection{Dataset}
The training dataset consists of 958 HDR images with varying resolutions, up to
4K.  None of the images contain absolute luminance values. The images are
scaled keeping their aspect ratio (and zero padded if necessary) to Full-HD
(1,920 $\times$ 1,080) resolution. The intensity range is randomly selected
during training, with maximum intensity chosen uniformly in the interval
[3,000, 5,000]. This random scaling works as a form of data augmentation and to
help prevent overfitting. The images are then clipped at the maximum display
intensity of 4,000 nits. The additional power-adaptation scalar is randomly
chosen using a uniform $\mathcal{U}[0,1]$ distribution for each mini-batch.
The test dataset used for evaluation is formed from 105 HDR images from the
Fairchild Photographic Survey~\cite{fairchild2007hdr}. These images contain
calibrated absolute luminance values and are not used during training.

\subsection{Optimisation}
The network was optimised until convergence of the loss for approximately 500,000
iterations, with $\beta = 20$. The Adam
optimiser~\cite{kingma2014adam} was used, with its default learning rate
$\lambda = 1e-3$ and $\beta_1 = 0.9$, $\beta_2 = 0.99$.  Training took 116 hours on an
NVIDIA RTX 2070 Super GPU using the PyTorch library~\cite{paszke2019pytorch}.

\section{Results}\label{sec:results}
This section presents results comparing DBLD with six other methods using quantitative analysis and qualitative visual inspection. In particular DBLD is compared against other methods:
Avg and Max~\cite{funamoto}, LP~\cite{cho2009backlight}, IMF~\cite{Liao2008Dynamic}, ZR \cite{zerman2015effects} and DM~\cite{narwaria2016dual}. 

\begin{figure*}[htb]
	\centering
	\includegraphics[width=1.0\linewidth]{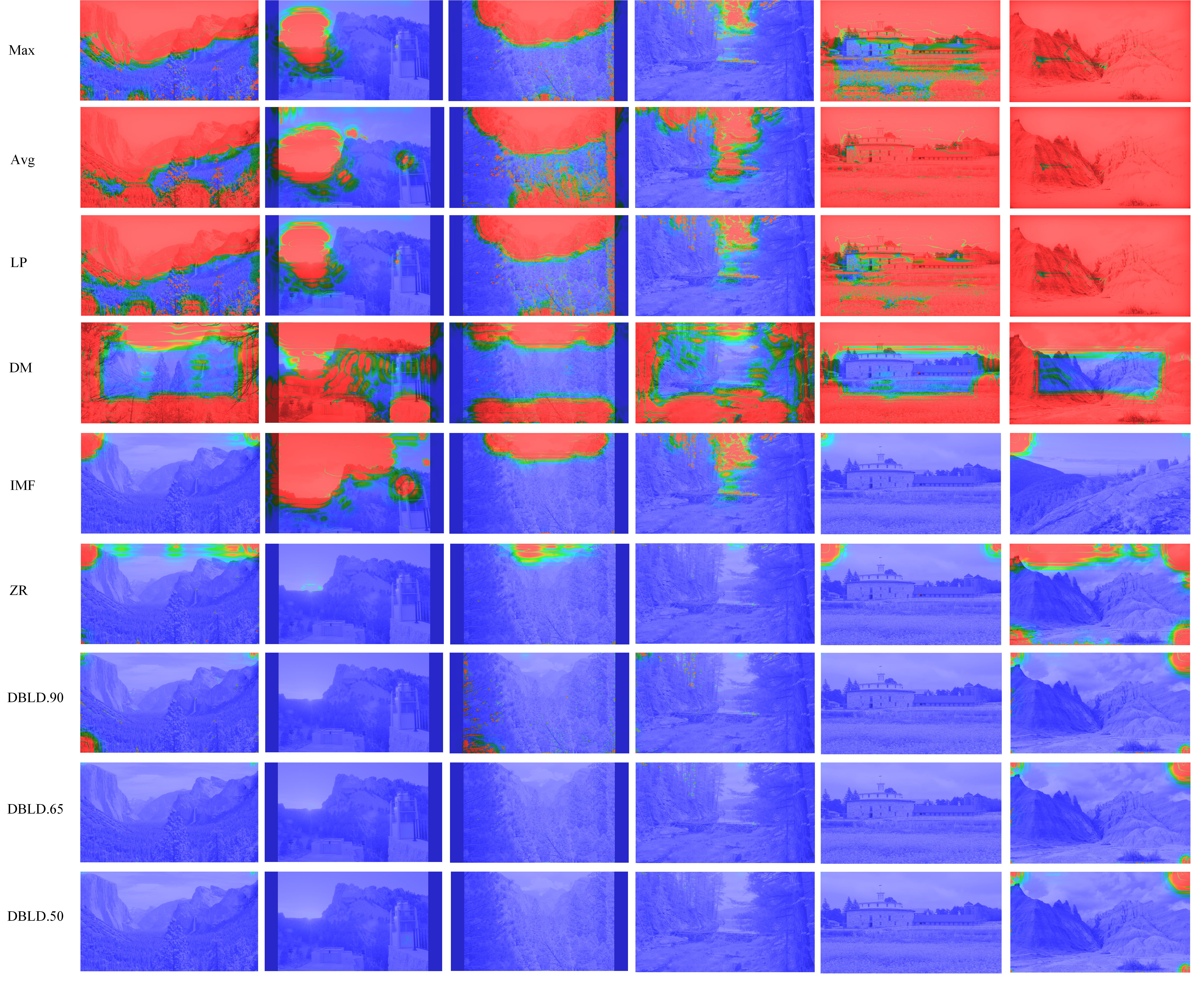}
    \caption{HDR-VDP-2.2 visibility probability maps for reconstructions of
    TunnelView(2), MtRushmore(1), KingsCanyon, AmikeusBeaverDamPM2,
HancockSeedField and BigfootPass using all methods. Blue indicates
imperceptible differences, red indicates perceptible
differences.}\label{hdrvdpmap}
\end{figure*}

\subsection{Quantitative evaluation}
\label{sec:quantitative}
DBLD is compared with the other methods using the evaluation
scheme proposed by Duan et al.~\cite{Duan}. The authors proposed computing
quantitative metrics using the reconstructed HDR images based on the model of
LCD described in Section~\ref{sec:hdr_reconstruction} and given by
equation~\ref{eq:displaymodel}. The authors demonstrated that there
is a strong correlation between objective and subjective evaluation of
different BLD algorithms~\cite{Duan}, making quantitative evaluation a viable
proxy to subjective experiments. A set of 105 HDR images from the Fairchild
Photographic Survey database were used to for the evaluation of the metrics. None of these 105 HDR images were
not used in the training of DBLD.

The metrics used for comparison were the Perceptually Uniform
(PU)~\cite{aydin2008extending} versions of PSNR, Multi-Scale
SSIM~\cite{wang2003multiscale}, along with
HDR\dash{VDP}\dash{2}.2~\cite{mantiuk2011hdr}. The power saving ratio (PSR)
\cite{burini2013modeling} corresponds to the percentage of power savings with
respect to the maximum display power, with higher values representing further
savings. 

Figure \ref{fig:plots} shows the results for the three quality metrics as a
function of power saving ratio.  For DBLD multiple values are computed by
adjusting \tparam and can be seen in Figure \ref{fig:plots} as points on the
curve. While DBLD was trained using \tparam values $\in [0, 1]$, results are
also shown for $\param > 1$ by extrapolation, demonstrating how the method
performs for very low power consumption. As can be seen, under most
circumstances, other methods fall under the curve demonstrating DBLD provides
better quality as a function of power usage.

Figure \ref{fig:violinplots} illustrates the distribution of results across the
105 tested images for all the methods and the three quality metrics as well as
the power saving ratio. As DBLD is adaptable to different outputs depending on
\tparam we show distributions with values of \tparam fixed to the values of 0.5
(DBLD.50), 0.65 (DBLD.65) and 0.9 (DBLD.90). These values of \tparam were
chosen to match the power consumption of state-of-the-art methods.  DBLD
outperforms all others except for ZR for PU-PSNR and HDR\dash{VDP}\dash{2}.2,
while for PU-MS-SSIM it achieves the first three positions.

\subsection{Visual inspection}
Figure ~\ref{hdrvdpmap} shows  the HDR-VDP-2.2 visibility probability maps for
all the methods for a selection of images from the testing dataset. The
HDR-VDP-2.2 visibility probability maps describe how likely it is for a
difference to be noticed by the average observer, at each pixel, between the
reconstructed HDR and the target HDR that is being displayed. Red values
indicate high probability, while blue values indicate low probability of
noticeable difference.

For DBLD, the same values of \tparam used in Section \ref{sec:quantitative} are
considered.  The results show that DBLD produces higher fidelity results than
the other methods and the number of perceivable artefacts reduces as \tparam
decreases.  In some methods, particularly the Avg, Max, LP and IMF methods,
brighter areas appear overexposed due to the low backlight values. The ZR
method can preserve more detail compared to these other methods.

\subsection{Timings}
DBLD takes an average of 0.061 seconds on an NVIDIA RTX 2070
Super GPU and 0.290 seconds on a (mobile) NVIDIA GTX 1050 Ti to render a Full-HD (1,920
$\times$ 1,080) image.  It is worth noting that these are not optimised timings,
using the model directly as implemented for training in Python.  Further
optimisations, for example rewriting code using a lower level language and
writing specialised kernels for the computational tree of the CNN can help to
further improve execution speed.

\section{Conclusion and future work}
In this work, a novel BLD method for HDR image rendering on HDR displays has
been proposed. The method uses a CNN to predict backlight values, trained on an
HDR image dataset. The method is also the first of its type to be controllable
and permits adjustment of power vs. quality. Objective evaluation of the method
is efficient and demonstrates improved image quality compared to other methods,
including current state\dash{of}\dash{the}\dash{art} algorithms.  Future work
will focus on further refinement of DBLD and extend it to process HDR videos
directly and in real\dash{time}.

\section*{Acknowledgment}
The authors would like to thank E. Zerman and M. Narwaria for providing source
code for their methods. Lvyin Duan also would like to thank the China
Scholarship Council (CSC) for their financial support.

\bibliography{bibi}

\end{document}